\documentclass[twocolumn,aps,prl,showpacs,superscriptaddress,
amsmath,amssymb]{revtex4}

\usepackage{color}
\usepackage{graphicx}
\usepackage{dcolumn}
\usepackage{bm}
\usepackage{textcomp}

\begin{document}

\title{Evidence of $d$-wave Superconductivity in
K$_{1-x}$Na$_{x}$Fe$_{2}$As$_{2}$ ($x = 0, 0.1$)
Single Crystals from
Low-Temperature Specific Heat Measurements}

\author{M.\ Abdel-Hafiez\footnote{
M.A.\ and V.G.\ have equally contributed to the present paper.}}
\affiliation{Leibniz-Institute for Solid State and Materials Research,
IFW-Dresden, D-01171 Dresden, Germany}
\author{V.\ Grinenko$^*$}\affiliation{Leibniz-Institute for Solid State and Materials Research, IFW-Dresden, D-01171 Dresden, Germany}
\author{S.\ Aswartham}\affiliation{Leibniz-Institute for Solid State and Materials Research, IFW-Dresden, D-01171 Dresden, Germany}
\author{{I.\ Morozov}\footnote{Present address:
Lomonosov Moscow State University, GSP-1, Leninskie Gory, Moscow, 119991, Russian Federation}}
\affiliation{Leibniz-Institute for Solid State and Materials Research, IFW-Dresden, D-01171 Dresden, Germany}
\author{M.\ Roslova$^\dagger$} 
\affiliation{Leibniz-Institute for Solid State and Materials Research, IFW-Dresden, D-01171 Dresden, Germany}
\author{O.~Vakaliuk}\affiliation{Leibniz-Institute for Solid State and Materials Research, IFW-Dresden, D-01171 Dresden, Germany}

\author{{S.\ Johnston} \footnote{Present address:
Dept.\ of Physics and Astronomy, University of Brit.\ Columbia, Vancouver, British Columbia, Canada V6T 1Z1}}
\affiliation{Leibniz-Institute for Solid State and Materials Research, IFW-Dresden, D-01171 Dresden, Germany}
\author{D.V.\ Efremov}
\affiliation{Leibniz-Institute for Solid State and Materials Research, IFW-Dresden, D-01171 Dresden, Germany}
\author{J.\ van den Brink}
\affiliation{Leibniz-Institute for Solid State and Materials Research, IFW-Dresden, D-01171 Dresden, Germany}
\affiliation{Institut f\"ur Theoretische Physik, TU Dresden, D-01062 Dresden, Germany}
\author{H.\ Rosner}
\affiliation{Max-Planck Institute for Chemical Physics of Solids, D-01187 Dresden, Germany}
\author{M.\ Kumar}\affiliation{Leibniz-Institute for Solid State and Materials Research, IFW-Dresden, D-01171 Dresden, Germany}
\author{C.\ Hess}\affiliation{Leibniz-Institute for Solid State and Materials Research, IFW-Dresden, D-01171 Dresden, Germany}
\author{S.\ Wurmehl}\affiliation{Leibniz-Institute for Solid State and Materials Research, IFW-Dresden, D-01171 Dresden, Germany}
\author{A.U.B.\ Wolter}\affiliation{Leibniz-Institute for Solid State and Materials Research, IFW-Dresden, D-01171 Dresden, Germany}
\author{B.\ B\"uchner}\affiliation{Leibniz-Institute for Solid State and Materials Research, IFW-Dresden, D-01171 Dresden, Germany}
\affiliation{
Institut f\"ur Festk\"orperphysik, TU Dresden, 
D-01062 Dresden, Germany}
\author{E.L.\ Green}\affiliation{Dresden High Magnetic Field Laboratory, Helmholtz-Zentrum Dresden-Rossendorf, 01314 Dresden, Germany}
\author{J. Wosnitza}\affiliation{Dresden High Magnetic Field Laboratory, Helmholtz-Zentrum Dresden-Rossendorf, 01314 Dresden, Germany}
\affiliation{Institut f\"ur Festk\"orperphysik, TU Dresden, D-01062 Dresden, Germany}
\author{P.\ Vogt}\affiliation{Kirchhoff-Institute for Physics, University of Heidelberg, D-69120 Heidelberg, Germany}
\author{A.\ Reifenberger}\affiliation{Kirchhoff-Institute for Physics, University of Heidelberg, D-69120 Heidelberg, Germany}
\author{C.\ Enss}\affiliation{Kirchhoff-Institute for Physics, University of Heidelberg, D-69120 Heidelberg, Germany}
\author{M.\ Hempel}\affiliation{Kirchhoff-Institute for Physics, University of Heidelberg, D-69120 Heidelberg, Germany}
\author{R.\ Klingeler}\affiliation{Kirchhoff-Institute for Physics, University of Heidelberg, D-69120 Heidelberg, Germany}
\author{S.-L.\ Drechsler} \email{s.l.drechsler@ifw-dresden.de}
\affiliation{Leibniz-Institute for Solid State and Materials Research, IFW-Dresden, D-01171 Dresden, Germany}

\date{\today}

\begin{abstract}
From the measurement and analysis of
the specific heat of high-quality
K$_{1-x}$Na$_{x}$Fe$_{2}$As$_{2}$ single crystals we establish the
presence of large  $T^{2}$ contributions with coefficients
$\alpha_{\rm sc}\approx 30$ mJ/mol K$^3$ at low-$T$
for
both $x$ = 0 and 0.1. Together with 
the observed
$\sqrt{B}$ behavior of the 
specific heat in the superconducting state both findings
evidence 
$d$-wave superconductivity on almost all
Fermi surface sheets with an average gap amplitude of
$\Delta_0$ in the range of 0.4 - 0.8 meV. The derived 
$\Delta_0$ and observed $T_c$ agree well
with the values calculated within 
Eliashberg theory, adopting a spin fluctuation mediated pairing 
in the intermediate coupling regime.
\end{abstract}


\maketitle
In spite of the substantial experimental and theoretical research efforts to
elucidate the symmetry and magnitude of the superconducting order parameter for
the Fe pnictides \cite{Hirschfeld,Stewart2011}, important questions
concerning the doping evolution of the superconducting gap remain 
unsolved \cite{Thomale,Maiti}.
In the stoichiometric parent
compounds
nesting usually occurs between electron (el) and hole (h) 
Fermi surface sheets (FSS) which is responsible for the presence of
long-range spin density wave (SDW) order. Superconductivity (SC)
emerges when the SDW order is suppressed by doping or external
pressure \cite{Stewart2011}. An $s_{\pm}$ gap symmetry (nodeless
gap function with opposite signs of the order parameters for el and
h pockets) is believed to be realized in under- and
optimally doped compounds,
since the
antiferromagnetic spin fluctuations (SF) on the vector $Q = (\pi,\pi)$ connecting the el and h pockets remain
strong in the vicinity of the SDW phase.
The situation in the overdoped regime is not so clear. With
further doping, el (h) bands disappear. Therefore, the paradigm of
the \textit{SF glue} at the vector $Q=(\pi,\pi)$
does not work. However,
SF have been found  
at some incommensurate propagation vectors
\cite{Lee2011}. This has led to several 
proposals for the order
parameters in heavily doped compounds:  extended $s$, $d$,
$s+id$ wave \cite{Thomale,Chubukov2012,Suzuki2011,Maiti2012}.
Thus, even from theoretical perspective the situation is
still controversial.

One of the most interesting families from this point of view is
Ba$_{1-x}$K$_x$Fe$_2$As$_2$.
Superconductivity in this compound occurs at 
$x \approx 0.15$ \cite{Avci2012}.
$T_c$ increases with K doping up to
$T_{\rm c} \approx 38$~K for $x$=0.4 (optimally doped regime). 
In this region experiments such as angle-resolved photoemission 
spectroscopy (ARPES) 
\cite{Evtushinsky2009} and thermal conductivity \cite{Reid1} 
show the absence of nodes in the gap, 
confirming an $s$-wave character of the order parameter.
The evolution of the SC order parameter at high 
hole-doping 
levels is not well studied, so far. The measurements on 
polycrystalline  samples show that
$T_c$ monotonically decreases with doping, reaching $T_c$= 3.5~K for
KFe$_2$As$_2$ (K122) \cite{Avci2012}. Whereas the el pockets are completely
gone in K122, it was theoretically proposed that a change of the 
order-parameter symmetry  to  $d$,  $s_{\pm}$ wave with accidental nodes, 
or $s+id$ should occur \cite{Thomale,Chubukov2012,Suzuki2011,Maiti2012}. 

Until now, there is no consensus in the interpretation of available experimental 
data in favor of 
one of the proposed order parameters. 
Indeed, measurements of thermal conductivity
\cite{Reid2,Dong} and of the London penetration depth \cite{Ha} have
been interpreted in terms of $d$-wave SC with line-nodes on each
FSS. In contrast, recent ultrahigh-resolution laser
ARPES data for K122 have been interpreted in terms of a rather
specific {\it partial}  nodal $s$-wave SC having an unusual gap with
\textgravedbl octet line-nodes\textacutedbl \ on the middle FSS, an almost-zero gap on the
outer FSS, and a nodeless gap on the inner FSS \cite{Okazaki2012}.
The specific-heat (SH) data for K122 above 0.4~K, only, were
interpreted within a weak-coupling BCS-like multiple SC gap scenario with line 
nodes
\cite{Fukazawa2009, Fukazawa2011}, adopting
multiband- $d$- or $s$- wave nodal SC.  But at the same time an extremely strong 
coupling and/or correlated regimes with heavy quasiparticles were suggested for 
the normal state. Moreover,
the authors of a subsequent SH study in magnetic fields \cite{Kim2011} 
doubted such an interpretation and denied the presence of a smaller second 
gap based on the observation of a non-intrinsic magnetic transition at low-$T$, 
probably due to unknown magnetic impurities. Naturally, under such circumstances 
a $T^2$ and/or a $\sqrt{B}$ behavior, generic for line-nodes in the 
SC gap 
\cite{Sigrist,Keubert1998,Dora2001}, could not be observed.  
Hence, further studies  
are necessary to clarify 
the symmetry of the SC order parameter and the magnitude of the coupling 
strength. These two issues are the main points of the present letter. 

K$_{1-x}$Na$_{x}$Fe$_{2}$As$_{2}$ single crystals with
$x$=0 [K122] and $x$=0.1
[(K,Na)122] were grown using KAs as a flux.
SH data in various fields were obtained by a relaxation technique in
a Physical Properties Measurement System (PPMS, Quantum Design).
The magnetic ac susceptibility as a function of temperature was also measured 
by use of 
the PPMS. The $T$-dependence
of the electrical resistivity was measured using a standard 4-probe
DC technique.
Two single crystals, K122 and (K,Na)122, were selected for SH measurements
below 0.4 K using the heat-pulse technique
within a dilution refrigerator.

\begin{figure}[b]
\includegraphics[width=18pc,clip]{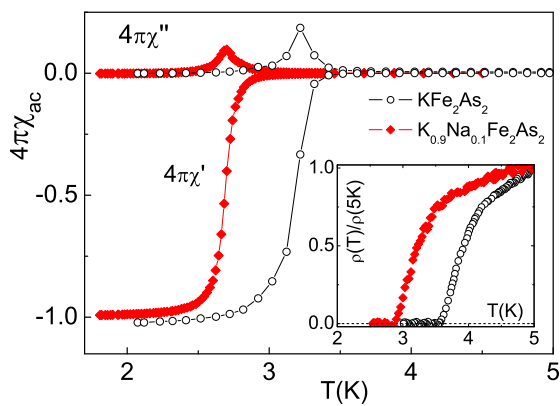}
\caption{(Color online) Temperature dependence of the complex volume ac susceptibility
$ \chi_{\rm ac} = \chi{'} + i\chi{''}$
measured with an amplitude of 5\,Oe and a frequency of $\nu$=
1\,kHz,
Inset: the normalized in-plane 
resistivity $\rho$ measured in zero-field.}
\label{Fig:1}
\end{figure}
\begin{figure}[t]
\includegraphics[width=18pc,clip]{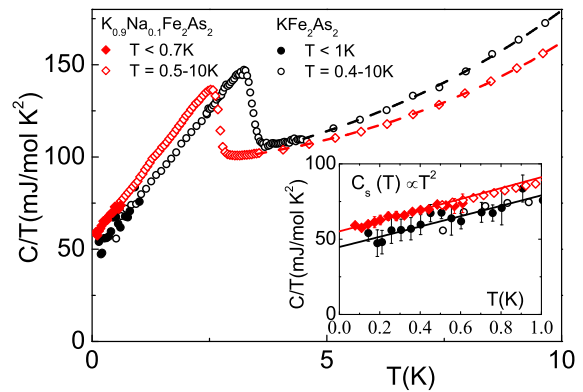}
\caption{(Color online) Specific heat, $C$, 
plotted as $C/T$ vs. $T$. Dashed lines: Fit 
in the normal state (see text).
Inset: zoom into the low-$T$ region, solid lines:
Fits using  Eq.~(\ref{eq2}). The data for K122 above 0.4~K are taken 
from Ref.\ \onlinecite{Abdel-Hafiez2012}.
} \label{Fig:2}
\end{figure}
\begin{figure}[b]
\includegraphics[width=18pc,clip]{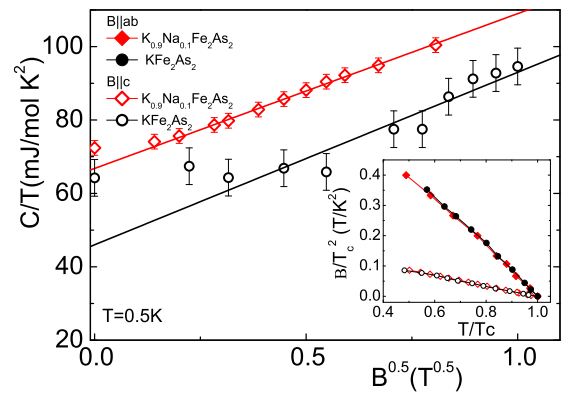}
\caption{(Color online) Field dependence of the specific heat, $C/T$, at 0.5~K. 
Solid lines:
Linear fits to define 
$P_{\rm SC}$ (see Eq. (\ref{eqp})). 
Inset: The scaled
upper critical fields for KFe$_{2}$As$_{2}$
\cite{Abdel-Hafiez2012} and K$_{0.9}$Na$_{0.1}$Fe$_{2}$As$_{2}$ 
vs.\ $T/T_{c}$.} \label{Fig:3}
\end{figure}
The $T$ dependence of the volume ac susceptibilities ($\chi{'}$
and $\chi{''}$) of K122 and (K,Na)122 are shown
in Fig.~\ref{Fig:1}. The sharp transition with  
4$\pi\chi{'}\sim$ 1
confirms the bulk nature of superconductivity and
the high quality of our crystals. The $T$-dependence of
the in-plane electrical resistivity evidences a drop to zero at 3.6 K
for K122 and at 2.9~K for (K,Na)122 in agreement with the
diamagnetic onset seen in the $\chi_{AC}$ data. The 
resistivity ratio 
$RR = \rho(300K)/\rho(5K)$
is found to be $\approx$ 120 for K(Na)122 with 
$\rho(5K)\approx 3.82 \mu\Omega$cm which is smaller than
$RR \approx$ 380 for K122 with $\rho(5K)\approx 1.44\ \mu\Omega$cm. Therefore, 
the reduced $T_c$ of (K,Na)122 can be attributed to 
the pair-breaking effects due 
to disorder induced by Na doping.

The zero-field specific-heat, $C$, of K122 and K(Na)122 
between 0.1 and 10 K is shown in Fig.\ \ref{Fig:2}.  
A clear sharp SC anomaly is observed at
$T_\mathrm{c}\approx$ 2.75~K for (K,Na)122 and at $T_c\approx$
3.5\,K for K122, in line with the resistivity and ac susceptibility 
data.
The jump height of $C$ at $T_c$ is
found to be $\Delta C/T_{c}$ $\approx$ 40~mJ/mol K$^2$ for
(K,Na)122. This value is a bit reduced from $\approx$
46~mJ/mol K$^2$ for K122. A zoom into the low-$T$
range is shown in the inset of
Fig.~\ref{Fig:2}. We observe a pronounced $T^2$-behavior in the SH for
 both crystals. Notice that {\it no} kinks and/or small hump 
anomalies, predicted 
for the low-$T$ region
by weakly coupled two-band approaches \cite{Fukazawa2009,Fukazawa2011}, 
have been observed in our measurements. The field dependence 
of $C/T$ at $T=0.5$~K obtained 
from SH measurements in magnetic fields \cite {supplement} are shown 
in Fig.\ \ref{Fig:3}. The observed $\sqrt{B}$ dependence 
together with the $T^2$ behavior provides a direct evidence for a
 $d$-wave SC order parameter
of K$_{1-x}$Na$_{x}$Fe$_{2}$As$_{2}$, at least for $x \leq 0.1$.

The $T$ dependence of the SH in the SC state allows one  
to determine whether the gap function possesses nodes.  For instance,  the 
exponential vanishing of $C$ as in 
conventional $s$-wave SC is caused by 
the finite gap in the quasiparticle spectrum. In contrast, in the
case of gap nodes quasiparticles are generated to the largest extent in
the vicinity of the gap nodes. In the case of line-nodes
the quasiparticle excitation spectrum
takes the form
$E_k=\hbar\sqrt{\upsilon_{\rm F}^{2}k_{\perp}^{2}+
v_\Delta^{2}k_{\parallel}^2}$
\cite{Durst2000,Vekhter2001},
where $k_{\perp}$ and $k_{\parallel}$ are
wavevectors perpendicular
and parallel to the FSS, respectively, $v_{\rm F}$ is the 
{\it renormalized} in-plane Fermi velocity 
at the position of the node, while 
$v_\Delta \approx \partial\Delta /\hbar \partial k $ is the
slope of the gap at the node associated with the dispersion of
the quasiparticles along the FS.
This leads to
a density of states (DOS)  linear in-energy. For the two-dimensional (2D) case 
it reads:
\begin{equation}\label{eqDOS}
N_{SC}(E) = \sum_{i}\frac{E}{\pi\hbar^2\upsilon_{\rm F}^i v_\Delta^i} ,
\end{equation}
where $i$ denotes the sum over all nodes. 
Here, we would like to note that both materials considered
are highly anisotropic, justifying the 2D approximation. The ratio of $H_{c2}$ 
(in-plane vs.\ out-off plane)
according to inset Fig.\ \ref{Fig:3} suggests a large mass 
ratio $\Gamma^2 \approx$ 25.
This value agrees with
DFT predictions for the ratio of the corresponding squared
plasma frequencies \cite{Abdel-Hafiez2012}.

Eq.\ (\ref{eqDOS}) results in a quadratic
power-law dependence $C_{s}\propto\alpha_{\rm sc}T^{2}$, with 
\begin{equation}\label{eqalfa}
\alpha_{\rm sc}\mbox{(mJ/mol K$^3$)} \approx
0.283\frac{\varkappa \gamma_{\rm el}\mbox{(mJ/mol K$^2$)}}
{\Delta_0\mbox{(meV)}}
\end{equation}
taken from Refs. \onlinecite{Keubert1998,Vekhter2001,Dora2001} 
and valid for SC 
if both the impurity scattering rate and 
magnetic field scales  are  smaller  than the temperature. 
Therefore, the observed $T^2$ behavior suggests that this condition is 
fulfilled for both crystals studied. Using
Eq.\ (\ref{eqDOS}) we also assumed that the energy gap at $T=0$, 
$\Delta_0$, is equal for all bands, while $\gamma_{el}$ 
is a {\it renormalized} normal-state Sommerfeld coefficient, which is an 
average 
value over all FSS (see the Supplement for details \cite{supplement}). 
In this context,
$\varkappa$ is a dimensionless factor taking into account the 
anisotropy of $1/v_{\rm F}$ at the position of the nodes with 
respect to the averaged 
isotropic
value. 
According to our preliminary {\it ab initio} calculations 
for the $k$-dependence of the DOS near the Fermi energy
 in the nodal direction, 
and the expected amount of the anisotropic mass renormalization
due-to anisotropic SF, $ \varkappa \sim 1.1$ to 1.15.
This range of values 
arises from the dominant contribution of the two inner FSS around
the $\Gamma$-point in the Brillouine zone, which is somewhat reduced
by a different antinodal anisotropy of the third FSS. 
The DOS of these
three FSS is
about 90\% to the total DOS.
The resulting anisotropy
is also in accord with the slightly larger gap value  obtained
in our Eliashberg-theory based calculations (see below) than the value 
derived from Eq.\ (\ref{eqalfa}).

To analyze the low-$T$ SH data we use the 
ansatz:
\begin{equation}
\label{eq2}
C= \gamma _{r}T + \alpha T^2+ \beta_{3}T^{3},
\end{equation}
where the cubic term $\beta_{3}$ comes from phonons, which 
is defined from the data in the normal state. 
The remaining terms stem from 
a magnetic cluster-glass contribution reported previously in 
Ref.\ \onlinecite{Grinenko2012} 
and from electronic degrees of freedom, which is the focus here. 
The best fit using Eq. (\ref{eq2}) at $T \leq$ 0.4K yields
$\gamma_{r}= 44.8$~mJ/mol K$^{2}$ and $\alpha_{\rm }= 33.7$~mJ/mol K$^{3}$ for
K122 and $\gamma_{r}=55.2$~mJ/mol K$^2$ and $\alpha_{\rm }= 35.4$~mJ/mol K$^{3}$
for (K,Na)122. It has been shown in Ref.\ 
\onlinecite{Grinenko2012} that the residual 
"Sommerfeld" coefficient $\gamma_r$ of K122 mainly stems 
from 
a cluster-glass phase. 
At variance with K122, for
(K,Na)122 an additional contribution 
to $\gamma_r$, 
exceeding
10~mJ/mol K$^{2}$,
caused by
a disorder-induced
pair-breaking effect \cite{Keubert1998} in 
accord with 
the enhanced residual resistivity of this crystal.
However, we emphasize that the $T^2$ contribution
measured by 
$\alpha$ is not an extrinsic \textgravedbl glassy\textacutedbl contribution but
stems from the generic electronic contribution in the 
superconducting state, only.
The estimated contribution of the cluster glass amounts
$\alpha_{\rm CG} \approx 2$~mJ/mol K$^{3}$ for K122 and 
$\alpha_{\rm CG } \approx 0$ for K(Na)122 \cite{Grinenko2012,supplement}.
Hence, we argue that our observed
large $T^{2}$ terms come mainly from 
quasiparticle excitations near line-nodes. The resulting coefficients are 
$\alpha_{\rm sc} \approx 32(3)$~mJ/mol K$^{3}$ for K122 and 
$\alpha_{\rm sc } \approx 35(1)$~mJ/mol K$^{3}$ for K(Na)122.
Such a large $T^2$ contribution to the low $T$ SH 
provides evidence of the $d$-wave character of the gap function.

In the case of a $d$-wave 
SC order parameter $C(B)/T=P_{\rm SC}\sqrt{B} + C(0)/T$ 
holds, if the energy associated with
the 
Doppler shift $E_H=
\hbar v_{\rm F}(\pi B/\Phi_0)^{0.5}$ is large 
as compared to 
$E_T=k_BT$ \cite{Vekhter2001}, where $\Phi_0$ is the flux quantum and 
$v_{\rm F} = 4\cdotp 10^{6}$ cm/s \cite{Abdel-Hafiez2012}. 
According to our data shown in Fig.\ \ref{Fig:3} 
for $T = 0.5$~K the field dependence of $C/T$ follows a $\sqrt{B}$-law 
at $B \gtrsim$ 0.06 $T$ 
which corresponds to $E_H/E_T\gtrsim$ 3 in accord with the 
theoretical prediction \cite{Vekhter2001}. 
Using the expressions (66, 67) in Ref. \onlinecite{Vekhter2001} for 
$P_{\rm SC}$ and $\alpha_{\rm sc}$ 
we get:
\begin{equation}\label{eqp}
\frac{P_{\rm SC}}{\alpha_{\rm sc}} \approx
\frac{\pi^{5/2}\hbar M_1v_{\rm F}}{54\zeta(3)k_B\Phi_0^{0.5}} 
\approx 1.8M_1\mbox{(K/T$^{0.5}$))},
\end{equation} 
where $M_1\approx$ 1 for a
liquid vortex state 
and $M_1 = 1/\pi^{0.5}\approx$ 0.56 for the 
disordered distribution \cite{Vekhter2001}. This gives $P_{\rm SC}$ in the range of 
30 - 60 mJ/mol K$^2$T$^{0.5}$ 
in accord
with 
the experimental 
values $ \approx 48$
~mJ/mol K$^{2}$T$^{0.5}$ 
for K122 and 
$ \approx 42$~mJ/mol K$^{2}$T$^{0.5}$ for (K,Na)122.    

Now, we will check 
whether the value of $\alpha_{\rm sc}$ can be satisfactorily described by
the standard $d$-wave mechanism on all FSS. 
Our analysis is based on Eq.\ (\ref{eqalfa}), which  was obtained
within a BCS-like theory. But it works very well also in the case of 
strongly coupled cuprates \cite{Ido1996}. Therefore, we expect its validity
for pnictides, too.
In addition,  we analyse 
the relevant coupling 
constants with the help of the strong-coupling  
Eliashberg equations.
The coupling constants 
$ \lambda_{\rm sf}$ of the el-SF and $\lambda_{\rm ph}$
of the electron-phonon interaction together with the two bosonic spectral 
densities
$\alpha^2F(\omega)$
determine the gap value and $T_c$ as well.
A broad spectral density for intraband SF
as obtained by recent
inelastic neutron scattering
measurements \cite{Lee2011}
can be described as:
 \begin{equation}
 \alpha^2F(\omega)=\frac{\lambda_{sf}\Gamma_{sf}}{\pi }\frac{\omega }{\omega^2 +\Gamma^2_{\rm sf}},
\label{eq7}
 \end{equation}
where $\Gamma_{\rm sf}=7.9$~meV.
For the phonons we assume a sharp Lorentzian spectral density with a
width of 0.5~meV centered at
$\omega_{\rm ph} \sim 20$~meV. Notice
that in our simple isotropic effective single-band
model considered here for a
$d$-wave superconducting order parameter, the el-ph coupling (EPC)
drops out from the gap equation by symmetry and it enters only  the
equation for the $Z$ function which
describes the bosonic mass renormalization of the h-like quasiparticles
\cite {supplement}.
Thus, here the EPC results in a slight {\it suppression} of $T_c$ and
$\Delta_0$ in contrast to the $s_{\pm}$ case, where the superconductivity
becomes stronger taking into account an intraband EPC.
The SC gap and $T_{\rm c}$ calculated within Eliashberg-theory and their
dependence on the SF coupling constant $\lambda_{\rm sf}$ for 
$\lambda_{ph}=0$ for the sake of simplicity
and also for the more realistic case, with 
a weak el-ph coupling constant $\lambda_{ph} \sim 0.2$ 
(according to density functional theory based 
calculations \cite{Abdel-Hafiez2012}) 
included,
are shown in
Fig.\ S2 of Ref.\ \onlinecite{supplement}.
In order to reproduce $T_{\rm c}$,
$\lambda_{\phi , {\rm  sf}} \sim 0.64=0.8\lambda_{\rm z, sf}$ should be adopted
for K122, where the former describes the strength of the paring interaction 
and the latter stands for the mass renormalization. 
(In the isotropic 
$s$-wave case $\lambda_{\phi , {\rm  sf}} = \lambda_{{\rm z,  sf}}$ holds.) 
These values for the coupling constants 
 together with the calculated EPC
constant $\lambda_{\rm ph}\approx 0.2$ \cite{Abdel-Hafiez2012} yield
a total coupling constant of $\lambda_{\rm z, tot} \approx 1$.
As a result we get  $\Delta (0)^{ET}\approx$0.75~meV  
, which is close to that 
 in BCS theory 
$\Delta_0^{BCS} \sim 0.65$~meV
obtained from the $2 \Delta_0/T_c$ ratio 
for a $d$-wave superconductor in the 2D case:
\begin{equation}
\frac{2\Delta_0^{BCS}  }{k_{\rm B}
T_{\rm c}}= \frac{4\pi}{\gamma_E \sqrt{e}} \approx  4.28  ,
\label{eqratio}
\end{equation}
where $\gamma_E$ is the Euler
constant $\gamma_E=1.781$.
The obtained $\lambda_{\rm z, tot}$ values suggest that we 
are
in the regime of 
intermediate coupling and we may apply Eq.\ (\ref{eqalfa}) to extract 
$\Delta_0$. 
Considering the isotropic case ($\varkappa = 1$) of Eq.(\ref{eqalfa}), 
the same gap 
function for all bands and using the experimental electronic linear coefficient 
$\gamma_{el} =$52-68~mJ/mol K$^2$ \cite{Grinenko2012}, we estimate 
$\Delta^{exp}_0$ 
in the range of 0.4 - 0.7 meV, which is close to the 
values obtained above theoretically. It is also
close to the larger gap values $\Delta_1 \approx 0.7$ 
to 0.8~meV obtained within effective 
weak-coupling two-band models for the electronically weakly connected bands  
\cite{Fukazawa2009,Fukazawa2011}.
However, the eight times smaller second gap $\Delta_2$ 
\cite{Fukazawa2009,Fukazawa2011} with a comparable partial DOS 
for the second effective band would produce 
too large $\alpha_{\rm sc}$ values exceeding our experimental value
by a 
factor of three or more.
Hence, such a multiband scenario can be excluded on the basis of our 
study \cite{remf}. 
In this context, we note that the SH jump 
$\Delta C/T_c\gamma_{el} \approx 0.7$ - 0.9 is close to the theoretical
weak coupling value of 0.95 for a $d$-wave superconductor \cite{Dora2001}.
Thus, the proposed effective single-band $d$-wave scenario  
is in a good agreement with experimental SH data.

Now, we compare the experimental value of $\alpha_{\rm sc}$ with that  
for the \textgravedbl octet line-nodes\textacutedbl 
scenario proposed in Ref.\ \cite{Okazaki2012}. In this 
case, the SH exhibits a
$T^2$ behavior at low $T$, too, however as shown in Ref.\  
\onlinecite{supplement}, the
experimental value of
$\alpha_{\rm sc}$ is
too large to be described by a
single gap with \textgravedbl octet line-nodes\textacutedbl 
on the middle ($\zeta$) FSS, only, as suggested in Ref.\
\onlinecite{Okazaki2012}.
Thus, we argue that the
recently proposed  interpocket $s_{\pm}$ scenario \cite{Maiti2012} with
accidental line-nodes on a particular FSS, only,
is rather unlikely to be realized in the bulk
as probed by the SH.
Therefore, despite  the multi-band topology of
K$_{1-x}$Na$_{x}$Fe$_{2}$As$_{2}$,
according to our SH data it behaves 
more or less 
like  a single-band
$d$-wave SC with corresponding line-nodes
on all FSS
as suggested 
by
Reid {\it et al.} based on
a thermal-conductivity study 
and universal scaling arguments valid for a
single-band
$d$-wave SC 
\cite{Reid1,Reid2}. It was 
predicted even before by theoretical calculations based on 
microscopic
weak-coupling theory \cite{Thomale}.
However, quantitatively, in the interpretation of the 
thermal conductivity in Refs.\ \onlinecite{Reid1,Reid2}
the nominal Sommerfeld coefficient 
$\gamma_{\rm n}$ was used and a possible magnetic
contribution was not subtracted.

In summary, high-quality 
K$_{1-x}$Na$_{x}$Fe$_{2}$As$_{2}$ (x=0,0.1) single crystals
were studied by specific-heat. The large $T^2$ contributions 
and the $\sqrt{B}$ behavior at low-$T$ evidence the presence of
line-nodes in the superconducting gap. From the experimental
data, an effective
gap amplitude $\Delta_0^{exp} \sim$~0.4~-~0.7~meV
was estimated. This $\Delta_0^{\rm exp}$ and $T_{\rm c}$ agree well 
with the gap value calculated
within Eliashberg-theory for a one-band $d$-wave superconductor 
implying a moderately strong electron-boson coupling 
constant $\lambda_{\rm z, tot} \approx 1$.
This
suggests that almost all, i.e.\  at least the three large Fermi-surface 
sheets 
with about 90\% of the total DOS (according to the LDA) have comparable
gap amplitudes. Our data provide 
the first 
direct and quantitative
evidence for 
$d$-wave superconductivity in a Fe-pnictide system. A 
detailed comparison with the cuprates might be helpful to deepen our
understanding of the paring mechanism
in both unconventional and still challenging superconducting families.

We thank
D.\ Evtushinsky and V.\ Zabolotnyy 
 for fruitful discussions 
as well as  M.\ Deutschmann, S.\ M\"{u}ller-Litvanyi, 
R.\ M\"{u}ller, J.\ Werner,
S.\ Pichl, K.\ Leger and S.\ Gass
for technical support. This
work was supported by the DFG through SPP 1458
and the E.-Noether program (WU 595/3-1
(S.W.)),
and the EU-Japan project (No. 283204 SUPER-IRON).
I.M.\, M.R.\ thank funding from RFBR (12-03-01143-a, 12-03-91674-ERA-a,  
12-03-31717 mol-a), and S.J.\ 
from FOM (The Netherlands), and S.W.\ thanks the
BMBF for support in the frame 
of the ERA.Net RUS project.

\vspace{3cm}
\widetext
\centerline{\bf \huge \hspace{9cm} SUPPLEMENT}
\vspace{1cm}
\centerline{
\makebox[13.5cm][c]{{\small  We present details of the
specific heat (SH) measurements 
in external magnetic fields and of the
 analysis of }}}
\centerline{
\makebox[13.5cm][c]{{\small data in the normal state. 
We also briefly sketch some aspects of the
renormalization of the normal-state Som- }}}
\centerline{
\makebox[13.5cm][c]{{\small merfeld coefficient
 and
details of our
Eliashberg-theory based calculations for $T_{\rm c}$ and the
superconducting or-}}}
\centerline{
\makebox[13.5cm][c]{{\small  der parameter $\Delta_0$ discussed in the main
text. For the clean limit case an estimate of the coefficient $\alpha_{\rm sc}$ which }}}
\centerline{
\makebox[13.5cm][c]{{\small 
 enters
 the
$T^2$ term in the low-$T$ SH expansion is provided for the
$s_{\pm}$-scenario with accidental line nodes pro-}}}
\centerline{
\makebox[13.5cm][c]{{\small  qposed recently, too.} }}

\vspace{-0.1cm}

\renewcommand{\theequation}{S\arabic{equation}}
\renewcommand{\thefigure}{S\arabic{figure}}
\renewcommand{\thetable}{S\arabic{table}}
\setcounter{equation}{0}
\setcounter{figure}{0}
\setcounter{table}{0}

\vspace{0.2cm}
\subsection{Specific-heat of K$_{0.9}$Na$_{0.1}$Fe$_{2}$As$_{2}$ (K(Na)122) in magnetic
fields}

\begin{figure}[b]
\includegraphics[width=40pc,clip]{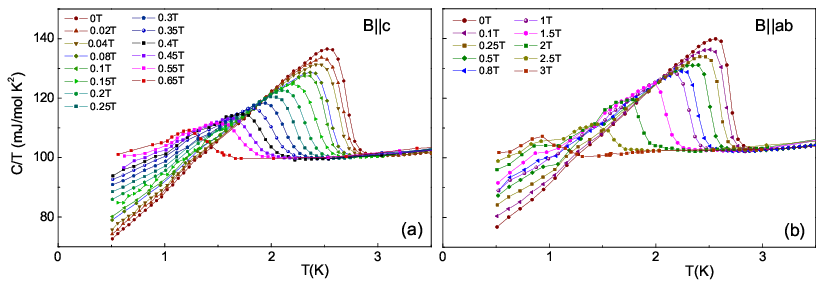}
\caption{Temperature dependence of the specific heat of
K$_{0.9}$Na$_{0.1}$Fe$_{2}$As$_{2}$ in various
applied magnetic fields parallel to the $c$ axis (a) and
parallel to the $ab$ plane (b).}
\label{SH}
\end{figure}

 The temperature dependence of
the SH of a
K(Na)122 single crystal in
magnetic fields $B\parallel c$ (a) and  $B\parallel ab$ (b)
is summarized in Fig.~\ref{SH}.
The SC
jump shifts and broadens systematically to lower temperature with increasing
field. The shift is more pronounced for $B \parallel c$ than for
$B \perp c$ reflecting the large upper critical field anisotropy
of this compound. The extracted temperature dependence of
the upper critical fields for both field orientations are
shown in the main text in Fig.~3. For the measurements in fields $B$
parallel $ab$
a small copper block was used to  orient
the sample. The SH of the copper block 
 was determined in
a separate measurement and subtracted from the data.
However, we observed a
small discrepancy between the data measured  
with that copper block (but subtracting its  contribution afterwards)
and without a copper block as
can be seen in Fig.~\ref{SH}, which originates in the
larger experimental error when using the
copper block.
Therefore, for the
quantitative analysis of the SH we used only data measured without
a copper block.
\vspace{1cm}

\subsection{Analysis of the specific-heat  in the normal
state}

To extract correctly 
$\alpha_{\rm sc}$ from the experimental data shown in Fig.\ 2 in
 the main text, at first, the normal state SH for both samples has been fitted
by the expression accounting for the
electronic, lattice, and magnetic
(due to a cluster glass (CG) phase) contributions as proposed in
Ref.\ \onlinecite{Grinenko2012s}:
\begin{equation}\label{eq1}
C_p = \gamma _{n}T + \varepsilon_{\rm CG}T^2 + \beta_{3}T^{3} + \beta_{5}T^{5},
\end{equation}
where $\gamma_{n}$= 95(2)~mJ/mol$\cdotp$K$^{2}$ for K(Na)122 and
89(3)~mJ/mol$\cdotp$K$^{2}$ for KFe$_{2}$As$_{2}$ (K122) are linear
contributions to the SH (denoted  as {\it
nominal} Sommerfeld coefficients) which are the sum of the standard
intrinsic electronic contributions $\gamma _{el}$  related to the
itinerant charge carriers (quasiparticles) (the so-called Sommerfeld
coefficient) and another linear, somewhat unusual glassy magnetic
contribution $\gamma_{CG}$ related to the CG-phase
\cite{Grinenko2012s}, and $\varepsilon_{\rm CG}$ is a quadratic
contribution due to the CG-phase. It was found
that $\varepsilon_{\rm CG}\approx 0$ for K(Na)122 and
$\varepsilon_{\rm CG}\approx 2$~mJ/mol$\cdotp$K$^3$ for K122
\cite{Grinenko2012s}. Finally, the lattice contributions are $\beta_3
= 0.556$~mJ/mol$\cdotp$K$^4$ and $\beta_5=1.16\cdot
10^{-3}$~mJ/mol$\cdotp$K$^6$\ for K(Na)122 and $\beta_3 =
0.589$~mJ/mol$\cdotp$K$^4$ and $\beta_5=1.2\cdot
10^{-3}$~mJ/mol$\cdotp$K$^6$ for K122.\\

Now, we briefly comment on the renormalization of the Sommerfeld coefficient 
due to the electron-boson interaction. 
The Eliashberg-theory, which we use to calculate $T_c$ and $\Delta$, 
considers the low-energy sector of characteristic bosonic energies. 
Within this approach  the relation between the SH of the 
"bare" electrons $\gamma^\eta_{el}$ and the renormalized quasiparticles 
due to 
electron-boson interaction $\gamma_{el}$ reads:
$$
\gamma_{\rm el}=\gamma^{\rm \eta}_{\rm el}(1+\lambda_{\rm z, tot} ),$$
where
$\lambda_{\rm z,tot} = \lambda_{z,sf }+\lambda_{z,ph}$ consists of two 
contributions: the electron-spin-fluctuation interaction $\lambda_{z,sf}$ 
and the electron-phonon interaction  $\lambda_{\rm z, ph}$ (see also the next subsection.) 
This renormalization gives a factor of about two for the enhancement of 
the Sommerfeld coefficient and similarly for the mass in the plasma 
frequency which determines the penetration depth and the 
superconducting 
condensate density for
$T\rightarrow 0$.
 The enhancement factor of the order $2.5-3$  of 
the Sommerfeld coefficient as compared to the LDA calculations 
$\gamma^{LDA}_{el} = 10.5$~mJ/mol K$^3$ and   $\gamma^\eta_{el} \sim 30$~mJ/mol K$^3$ is due to 
a high-energy renormalization  of the bands by self-energy effects 
beyond the LDA based approach 
seen also in the ARPES data and similarly in the optical mass which enters
the Drude weight, i.e., the intraband plasma frequency at high temperature
and energies as compared to the energies of the low-energy bosons included in our 
Eliashberg-theory based approach
(for details see Refs.\ S2-S5). 
   
\vskip 0.5cm

\subsection*{Details for the $d$-wave Eliashberg-theory based calculations}
We consider here, for the sake of simplicity, a two-dimensional (2D) single-band
$d$-wave superconductor
with a cylindrical Fermi surface parameterized by the angle $\theta$.
In general,
 the momentum dependence of the electron-boson spectral function
for a given mode can be expanded in terms of Fermi surface harmonics
\begin{equation}
\alpha^2F(\omega,\theta,\theta^\prime) =
\sum_{JJ^\prime} Y_J(\theta) \alpha^2F(\omega,J,J^\prime) Y_{J^\prime}(\theta'),
\end{equation}
where the functions $Y_{J}$ denote the spherical harmonics.
For simplicity, we assume that $\alpha^2F(\omega,\theta,\theta^\prime)$
is dominated by contributions
from $J = 0$ ($Y_0 \equiv Y_z = 1$) and $J = 2$
($Y_2 = Y_\phi \equiv \sqrt{2} \cos(2\theta)$)
and that it is diagonal in the basis $J$, $J^\prime$.
This is the minimal set required to model $d$-wave superconductivity.
Then the electron-boson spectral function can be parameterized via dimensionless 
coupling constants $\lambda_{z}$ and $\lambda_\phi$:
\begin{equation}
    \alpha^2F(\nu,\theta,\theta ')=\sum_i [\lambda_{i,z} + 2 \lambda_{i,\phi}\cos(2\theta)\cos(2\theta^\prime)]F_i(\nu) \quad ,
\end{equation}
where $i$ is a mode index,  
$F_i(\omega) = B_i(\omega)/\left[2 \int_0^\infty B_i(\nu)d\nu/\nu\right] $ 
is a
normalized electron-boson spectral function, and $B_i(\omega)$ is 
the
bosonic density of states. 
We will consider contributions from two types -
a single dispersionless 
optical-phonon ($i=$~ph) branch and a spectrum of antiferromagnetic
spin fluctuations ($i=$~sf).

It is useful to introduce:
\begin{equation}
    \lambda_{z,\phi}(n-m)
    = \lambda_{z,\phi} \sum_i \int_0^\infty \frac{2\nu F_i(\nu) d\nu}{\nu^2 + (\omega_n - \omega_m)^2 } \quad ,
\end{equation}
where $\omega_n$ and $\omega_m$ are fermion-Matsubara frequencies. 
The set of the imaginary-axis Eliashberg-equations for a $d$-wave
superconductor are then:
\begin{equation}
    \phi(i\omega_n) = \frac{\pi}{\beta}\sum_{m} 2\lambda_\phi(m-n) \left\langle
    \frac{\phi(i\omega_m)\cos^2(2\theta)}{\sqrt{\omega^2_mZ^2(i\omega_m) + \phi^2(i\omega_m)\cos^2(2\theta)}}
    \right\rangle_\theta \quad ,
\end{equation}
 and
\begin{equation}
\omega_nZ(i\omega_n) = \omega_n + \frac{\pi}{\beta}\sum_{m} \lambda_z(m-n) \left\langle
    \frac{\omega_mZ(i\omega_m)}{\sqrt{\omega^2_mZ^2(i\omega_m) + \phi^2(i\omega_m)\cos^2(2\theta)}}
    \right\rangle_\theta \quad ,
\end{equation}
where
$\beta = 1/{\rm k}_{\rm B}T$ is the inverse temperature,  
$\langle \dots \rangle_\theta$ denotes a Fermi surface average, and 
the Matsubara frequency sum is cut-off at $m = 500$.  
The gap function is then given by $\Delta(i\omega_n) = \phi(i\omega_n)/Z(i\omega_n)$.
The transition temperature $T_{\rm c}$ is determined from the highest temperature at which
$\phi$ has a non-zero solution while the gap magnitude $\Delta$($T$= 0) is
approximated with a low-temperature ($T\sim 0.2$~K) value of
$\Delta(i\omega_0 = i\pi/\beta)$.


To describe the spin fluctuations we adopt the usual form for the bosonic 
density of states
$B_{sf}(\nu) = \Gamma_{sf}\frac{\nu}{\nu^2 + \Gamma^2_{sf}} \ ,$
with $\Gamma_{sf} = 7.9$ meV.
The high-frequency tail of this spectrum was cutted at 100~meV.
For the optical-phonons we assume a narrow Lorentzian line-shape
$B_{ph}(\nu) = \frac{\Gamma_{ph}}{(\nu-\Omega_{ph})^2 + \Gamma_{ph}^2} - \frac{\Gamma_{ph}}{(\nu+\Omega_{ph})^2 + \Gamma_{ph}^2} \quad ,$
centered at
$\Omega_{ph} = 20$ meV and  $\Gamma_{ph} = 0.5$ meV.  Furthermore, 
the spectrum for the spin-fluctations is cut-off at 100 meV. 
We further assume that the phonons do not contribute to the
$d$-wave pairing ($\lambda_{\phi, ph} = 0$) and we adopt the so called 
unbalanced
scenario for $d$-wave supercondutivity [S7], $ \lambda_{\phi, sf} =0.8 \lambda_{\rm z, sf} $. This ratio  is also in  good agreement with 
$\lambda_{s,d}$ in Ref.\ \onlinecite{s6}.

\begin{figure}
\includegraphics[width=\textwidth]{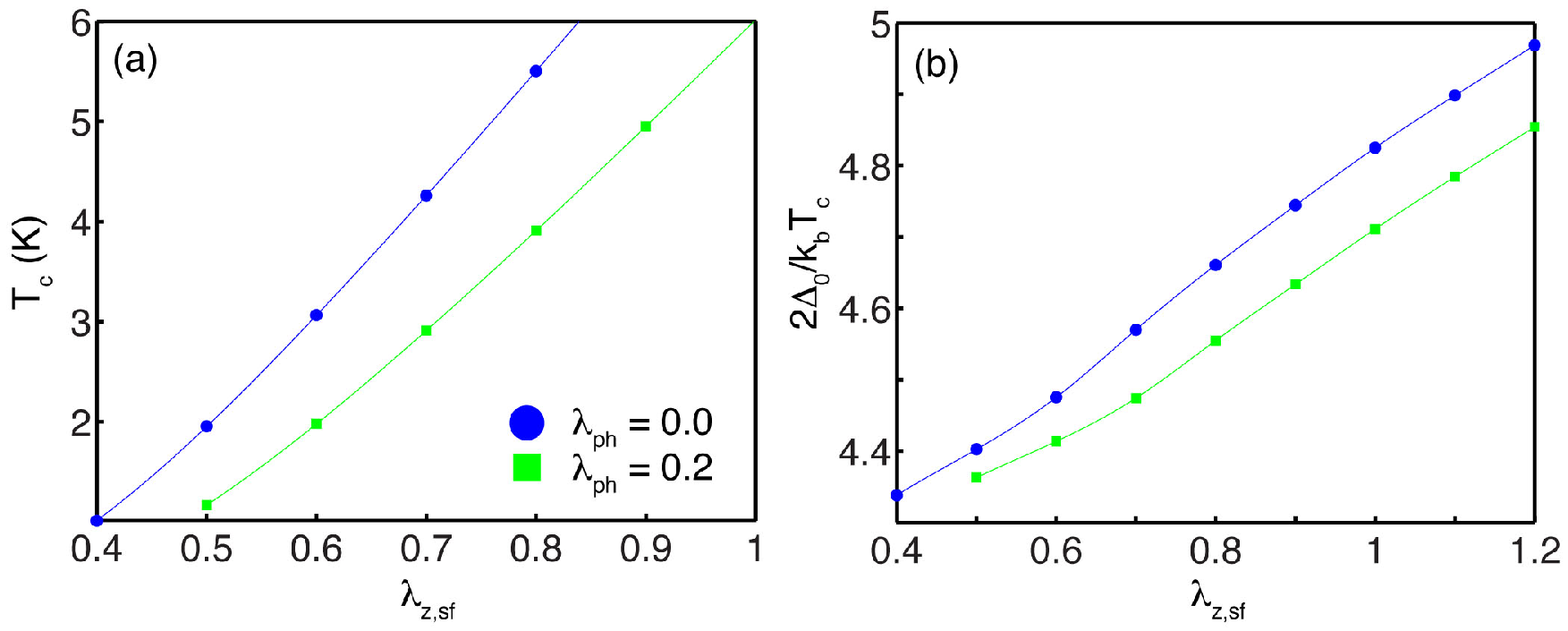}
 \caption{\label{Fig:deltaTc}
 (a) 
 Superconducting transition temperature $T_{\rm c}$ obtained from
 our model calculations as a function of $\lambda_{z,sf}$.
 (b) The same as in (a) for the corresponding gap to $T_c$
 ratio $2\Delta/{\rm k}_{\rm B}T_{\rm c}$.
 }
\end{figure}

In Fig.\ \ref{Fig:deltaTc} we plot  $2\Delta(T$=$0)/{\rm k}_{\rm B}T_{\rm c}$
extracted by use of
this model.
For $\lambda_{\rm sf}$ and $\lambda_{\rm ph}=0$,
this ratio
tends to about
4.28 
(not shown in Fig.\ S2). Thus, the electron-phonon
coupling reduces the strong-coupling corrections in this special case.

\subsection{Comparison
of $\alpha_{\rm sc}$ for $d-$ and
$s_{\pm}$-wave (with accidental nodes) SC order parameters }

The experimental value of $\alpha_{\rm sc}$ discussed in the main
text is too large to be described by a gap with "octet-line nodes"
on the middle ($\zeta$) FSS, only, as suggested in Ref.\
\onlinecite{Okazaki2012s}. In fact, let us consider the ratio between
$\alpha_{\rm sc}^d$ for a $d$-wave gap with 16 nodes on all FSS
having nearly equal gaps $\Delta_0
\approx$~0.6meV and $\alpha_{\rm sc}^s$ for a single $s$-wave gap
with 8 nodes on a particular FSS. Using the
general expression for the coefficient $\alpha_{\rm sc}$ in the $T^2$ term
entering 
Eqs.\ (1) and (2) in the main
text, we have:
\begin{equation}\label{s_eqDOS}
N_{SC}(E) = \sum_{i}\frac{E}{\pi\hbar\upsilon_{\rm F}^i\partial\Delta^i/\partial k}
\approx \frac{E}{\partial\Delta/\partial\theta}\sum_{i}\varkappa^iN^i(\varepsilon_{\rm F}).
\end{equation}
Here, we assumed that $\partial\Delta/\partial k =
\partial\Delta/\partial\theta\frac{1}{k_{\rm F}}$ and then 
$N^i(\varepsilon_{\rm F})$ is the DOS in the normal state corresponding to
the i$^{th}$ FSS with a node. According to
the considerations given in 
the main text we have:
\begin{equation}\label{s_eqalfa}
\alpha_{\rm sc} \approx
\frac{9}{2}\zeta (3)k_{\rm B}^3\frac{E}{\partial\Delta/\partial\theta}\sum_{i}\varkappa^iN^i(\varepsilon_{\rm F})
\approx
0.283\frac{\varkappa \gamma_{\rm el}}
{\Delta_0}.
\end{equation}
The last expression is written for the case of a
$d$-wave SC gap ($\partial\Delta/\partial\theta = 2\Delta_0$)
measured in meV and
where $\zeta$ denotes Riemann's zeta function and
$\zeta(3)=1.2$. We used also that the
bare Sommerfeld coefficient
$\gamma_{\rm el} =\frac{\pi^2k_B^2}{3}N_{tot}(\varepsilon_{\rm F})$.
For the sake of simplicity, we also assumed that all
FSS have similar $\varkappa$ values, therefore, 
$\sum_{i}\varkappa^iN^i(\varepsilon_{\rm F})=4\varkappa
N^i(\varepsilon_{\rm F}) =\varkappa N_{{\rm tot}}(0)$.
Finally, from Eq.\ (\ref{s_eqalfa}) for
$T\ll \Delta_s^{min}\approx$~1.3~K \cite{Okazaki2012s} we
estimate:
\begin{equation}
\frac{\alpha_{\rm sc}^d}{\alpha_{\rm sc}^s}
\approx
\frac{\varkappa\sum_{i}N^i(\epsilon_{\rm F})(1+\lambda^{i})/\partial\Delta_d/\partial\theta}
{2\varkappa^{\zeta}N^{\zeta}(\epsilon_{\rm F})(1+\lambda^{\zeta})
/\partial\Delta_s/\partial\theta } \approx 4.5,
\end{equation}
where we used (according to our calculations and Ref.\ \cite{Has}) that
$ \frac{\varkappa \sum_{i}N^i(\epsilon_{\rm F})(1+\lambda^{i})}
{\varkappa^{\zeta} N^{\zeta}(\epsilon_{\rm F})(1+\lambda^{\zeta})}
\sim 5
$
with $\lambda^{\zeta}\approx\lambda^{tot}$,
$\varkappa$/$\varkappa^{\zeta} \sim 1.3$, and
$\partial\Delta_s/\partial\theta\approx
1.8\partial\Delta_d/\partial\theta$, where $\varkappa^{\zeta}$ and
$\partial\Delta_s/\partial\theta$ have been calculated at the nodal directions 
of
the $\zeta$ FSS according to the
data given in Ref.\ \onlinecite{Okazaki2012s}. For the physical 
consequences of this estimated ratio
with respect to the symmetry of the superconducting order 
parameter, see the main text.

\end{document}